\documentclass[aps,prl,twocolumn,superscriptaddress]{revtex4-2}
\usepackage{amsmath}
\usepackage{mathtools} 
\usepackage{amssymb}
\usepackage{amsthm}
\usepackage{bm} 
\usepackage{upgreek} 
\usepackage{xcolor}
\usepackage{graphicx}
\usepackage{epstopdf}
\usepackage{ulem}
\usepackage[colorlinks,linkcolor=blue,anchorcolor=blue,citecolor=blue,urlcolor=blue]{hyperref}
\usepackage{braket}
\usepackage{orcidlink}

\begin{document}
\title{Theory of Pressure Dependence of Superconductivity in Bilayer Nickelate La$_3$Ni$_2$O$_{7}$}

\author{Kai-Yue Jiang\orcidlink{0009-0007-0395-9662}} \thanks{K.Y.J, Y.H.C, and Q.G.Y. contributed equally to this work.}
\affiliation{School of Physics and Physical Engineering, Qufu Normal University, Qufu 273165, China}

\author{Yu-Han Cao} \thanks{K.Y.J, Y.H.C, and Q.G.Y. contributed equally to this work.}
\affiliation{National Laboratory of Solid State Microstructures $\&$ School of Physics, Nanjing University, Nanjing 210093, China}

\author{Qing-Geng Yang} \thanks{K.Y.J, Y.H.C, and Q.G.Y. contributed equally to this work.}
\affiliation{National Laboratory of Solid State Microstructures $\&$ School of Physics, Nanjing University, Nanjing 210093, China}

\author{Hong-Yan Lu\orcidlink{0000-0003-4715-7489}} \email{hylu@qfnu.edu.cn}
\affiliation{School of Physics and Physical Engineering, Qufu Normal University, Qufu 273165, China}

\author{Qiang-Hua Wang\orcidlink{0000-0003-2329-0306}} \email{qhwang@nju.edu.cn}
\affiliation{National Laboratory of Solid State Microstructures $\&$ School of Physics, Nanjing University, Nanjing 210093, China}
\affiliation{Collaborative Innovation Center of Advanced Microstructures, Nanjing University, Nanjing 210093, China}

\begin{abstract}
The recent experiment shows the superconducting transition temperature in the Ruddlesden-Popper bilayer La$_3$Ni$_2$O$_{7}$ decreases monotonically with increasing pressure above 14 GPa. In order to unravel the underlying mechanism for this unusual dependence, we performed theoretical investigations by combining the density functional theory (DFT) and the unbiased functional renormalization group (FRG). Our DFT calculations show that the Fermi pockets are essentially unchanged with increasing pressure (above 14 GPa), but the bandwidth is enlarged, and particularly the interlayer hopping integral between the nickel $3d_{3z^2-r^2}$ orbitals is enhanced. From the DFT band structure, we construct the bilayer tight-binding model in terms of the nickel $3d_{3z^2-r^2}$ and $3d_{x^2-y^2}$ orbitals. On this basis, we investigate the superconductivity induced by correlation effects by FRG calculations. We find consistently $s_\pm$-wave pairing triggered by spin fluctuations, but the latter are weakened by pressure and lead to a decreasing transition temperature versus pressure, in qualitatively agreement with the experiment. We emphasize that the itinerancy of the $d$-orbitals is important and captured naturally in our FRG calculations, and we argue that the unusual pressure dependence would be unnatural, if not impossible, in the otherwise local-moment picture of the nickel $d$-orbitals. This sheds lights on the pertinent microscopic description of, and more importantly the mechanism of superconductivity in La$_3$Ni$_2$O$_{7}$. 
\end{abstract}
\maketitle

{\it Introduction}. Following the discovery of cuprate and iron-based high-temperature superconductors, the signature of superconductivity in the bilayer Ruddlesden-Popper (RP) La$_3$Ni$_2$O$_{7}$ (La327), with a critical temperature $T_{c}$ near 80 K above 14 GPa \cite{2023Nature}, has being attracting widespread research interest, both experimentally  \cite{exp-9,exp-13,exp-16,exp-28,exp-37,exp-38,exp-41,exp-43,exp-44,exp-48,exp-49,exp-50,exp-51,exp-53,exp-54,exp-56,exp-57,exp-61,exp-68,exp-69,total} and theoretically \cite{t-1,t-2,t-3,t-5,t-6,t-7,t-8,t-60,t-10,t-11,t-12,t-14,t-15,t-17,t-18,t-19,t-21,t-22,t-4,t-20,t-23,t-36,t-47,t-59,t-24,t-25,t-26,t-27,t-58,t-39,t-42,t-45,327arxivhu2,t-46,t-55,t-58,t-60,t-62,t-63,t-64,t-65,t-66,t-67,t-70,t-72,total}. La327 features a NiO$_{2}$ plane, akin to the CuO$_{2}$ plane in cuprates, exhibiting quasi-two-dimensionality. The average valence states of nickel (Ni) atoms in La327 is Ni$^{2.5+}$ ($3d^{7.5}$), with the mixed valence states between $3d^{8}$ and $3d^{7}$. The $3d_{3z^2-r^2}$ and $3d_{x^2-y^2}$  orbitals are approximately half-filled and quarter filled, respectively, and are active near the Fermi level ($E_\mathrm{F}$). In view of the three Fermi pockets and the multi-orbital nature, La327 is closer to the situation in iron pnictides. In comparison, in the infinite-layer nickelates, the valence state is Ni/Cu$^{1+}$ ($3d^{9}$) and is closer to cuprates \cite{infinite-layer_nickelate,Anderson1987,Zhang-Rice_1988}. The theories proposed for La327 so far suggest that the strong vertical interlayer coupling between the Ni $3d_{3z^2-r^2}$ orbitals is important in driving Cooper pairing, with $s_\pm$-wave pairing symmetry  \cite{t-2,t-5,t-6,t-14,t-15,t-17,t-21,t-24,t-25,t-26,t-45,327arxivhu2}, or  $d$-wave pairing symmetry \cite{t-4,t-23,t-20,t-59,t-36,t-47,t-55,t-58,t-67}. 

Very recently, a comprehensive high pressure experiment up to 104 GPa is performed for the bilayer RP La327 single crystals \cite{327pressure}. From 0 to 14 GPa, La327 exhibits the $Amam$ space group and remains weakly insulating. Above 14 GPa, it undergoes a phase transition to the $I4/mmm$ structure and enters the superconducting state. The transition temperature decreases monotonically from 83 K at 18 GPa, where the superconductivity just emerges with pressure, and drops to below 40 K for pressures above 90 GPa, forming a right-triangle shape in the temperature-pressure phase diagram. 
We argue that this pressure dependence is unusual. First, we can rule out the role of electron-phonon coupling, since the transition temperature is well above the McMillan limit ($\sim$ 40 K) \cite{MMlimit}. On the other hand, the pressure dependence should be inconsistent with the local-moment picture of the electrons, such as the multi-orbital $t$-$J$ model. In such a picuture, a rough estimate of the transition temperature is $T_{c} \sim J \exp[-\frac{1}{N_F J}]$, where $J$ is the effective spin-exchange and $N_F$ is the density of states (DOS) at the $E_\mathrm{F}$. We observe that $J \propto t^2$ and $N_F\propto 1/t$, where $t$ is a representative hopping integral. Since $t$ should increase under hydro-static pressure, we see that both $J$ and $N_F J$ would increase with pressure, yielding an increasing $T_c$, which is apparently at odd with the experiment \cite{327pressure}. It is therefore important to ask whether the unusual pressure dependence of $T_c$ in La327 could be understood in the itinearant picture. In turn, the experimental pressure dependence can be used to judge the relevance of the itinerant picture and subsequently unravel the mechanism of superconductivity.  
\begin{figure*}[htbp]
	\centering
	\includegraphics[width=1\linewidth]{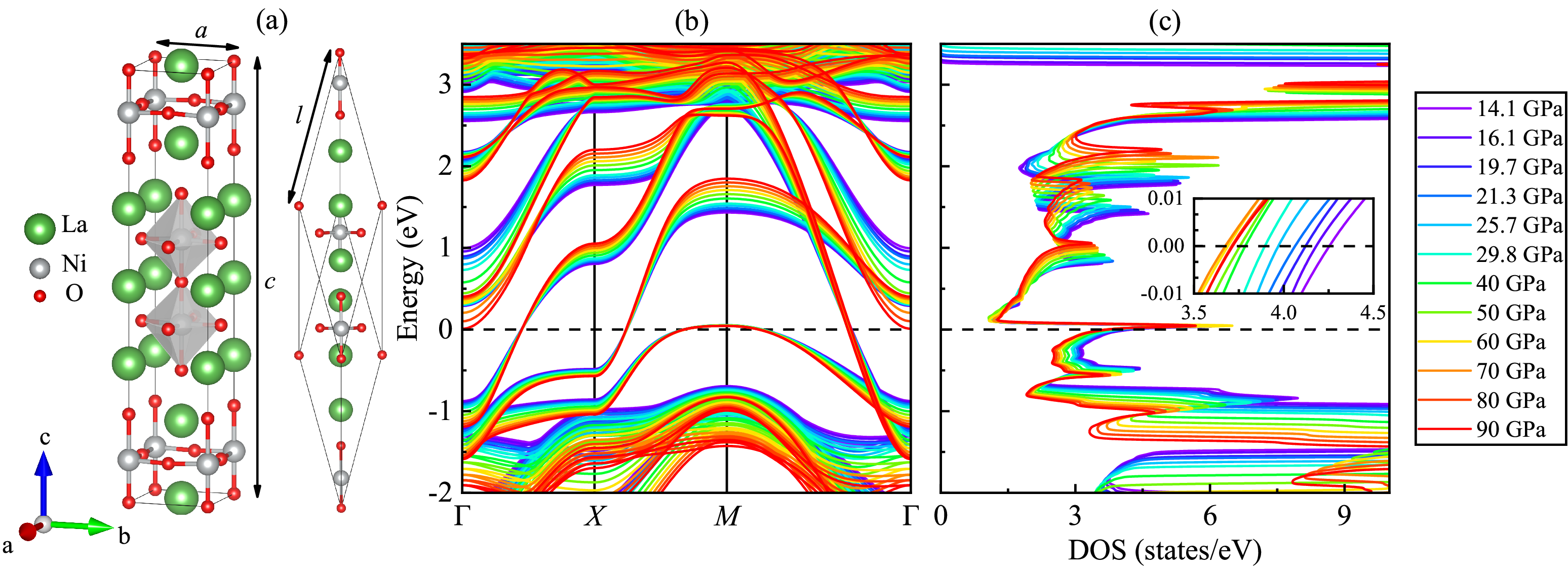}
	\caption{(a) Conventional cell and primitive cell for La327 under high pressure. (b) Band structure and (c) DOS under different pressures. The subfigure shows the DOS near the $E_\mathrm{F}$.}
	\label{F1}
\end{figure*}

In this Letter, we systematically study the electronic structures of La327 in the expreimental pressure regime by density functional theory, construct the bilayer two-orbital tight-binding model in terms of the $3d_{3z^2-r^2}$ and $3d_{x^2-y^2}$ orbitals of Ni, and then analyze the correlation effects and superconductivity by the unbiased singular-mode functional renormalization group (SM-FRG). We demonstrate that the pairing symmetry is consistently $s_\pm$-wave, and the pairing is triggered by spin fluctuations, which are weakened by increasing pressure, leading to monotonic decrease of the transition temperature, in qualitative agreement with the experiment. The results support the itinerant picture of the pairing in the bilayer La327 under high pressures. 

{\it Electronic model}.
The lattice structure of La327 under high-pressure with $I4/mmm$ space group is shown in Fig.~\ref{F1}(a). The electronic structures are performed by DFT calculations as implemented in Vienna {\it ab initio} simulation package (VASP) \cite{vasp}. The projector augmented-wave (PAW) \cite{PAW} method and the generalized gradient approximation (GGA) of Perdew-Burke-Ernzerhof (PBE) \cite{PBE} exchange-correlation functional are adopted. The cutoff energy of plane-wave is set as 500 eV. The $\Gamma$-centered $12\times12\times12$ and $36\times36\times36$ $k$-point grids for the self-consistent and DOS calculations are used, respectively. Based on the experimental lattice parameters listed in Table~\ref{T1}, fully optimization is carried out on the lattice parameters and atomic positions. The optimized lattice parameters are also shown in Table~\ref{T1}. 
\begin{table}[htbp]
	\caption{The lattice parameters (\AA) under different pressures (GPa) from experiment (exp.) and DFT calculation (cal.), where $a$ and $c$ ($l$) represent the conventional cell (primitive cell) lattice parameters.}
	\centering
	\setlength{\tabcolsep}{5pt}
	\renewcommand{\arraystretch}{1.5}
	\begin{tabular}{ccccccc}
		\hline
		\hline
		Pressure & $a_\mathrm{exp.}$ & $c_\mathrm{exp.}$ & $l_\mathrm{exp.}$ & $a_\mathrm{cal.}$ & $c_\mathrm{cal.}$ & $l_\mathrm{cal.}$   \\ \hline
		14.1  & 3.756 & 20.020  & 10.356  & 3.754 & 19.788  & 10.244  \\ 
		16.1  & 3.747 & 19.967  & 10.329  & 3.743 & 19.724  & 10.211  \\ 
		19.7  & 3.732 & 19.833  & 10.262  & 3.725 & 19.628  & 10.161  \\ 
		21.3  & 3.726 & 19.746  & 10.218  & 3.717 & 19.583  & 10.138  \\ 
		25.7  & 3.710 & 19.648  & 10.168  & 3.698 & 19.471  & 10.080  \\ 
		29.8  & 3.686 & 19.591 & 10.136 & 3.679 & 19.375  & 10.031  \\ 
		40.0  & - & - & - & 3.640 & 19.159  & 9.919  \\ 
		50.0  & - & - & - & 3.605 & 18.976  & 9.825  \\ 
		60.0  & - & - & - & 3.574 & 18.809  & 9.738  \\ 
		70.0  & - & - & - & 3.546 & 18.663  & 9.662  \\ 
		80.0  & - & - & - & 3.520 & 18.527  & 9.592  \\ 
		90.0  & - & - & - & 3.450 & 18.402 & 9.527 \\ 
		\hline
		\hline
		\label{T1}
	\end{tabular}
\end{table}
\begin{table*}[htbp]
	\caption{On-site energies $\varepsilon_a$ and hopping integrals $t_\delta^{ab}$ of the bilayer two-orbital tight-binding model for La327 under different pressures. Here, $x$/$z$ denotes the $3d_{x^2-y^2}$/$3d_{3z^2-r^2}$ orbital. Note that the vertical interlayer distance is assigned as $\frac12$. The unit of pressure is GPa, and the unit of $\varepsilon_a$ and $t_\delta^{ab}$ are eV.}
	\centering
	\setlength{\tabcolsep}{10pt}
	\renewcommand{\arraystretch}{1.5}
	\begin{tabular}{ccccccccccc}
		\hline
		\hline
		Pressure & $\varepsilon_x$ & $\varepsilon_z$ & $t_{(100)}^{x x}$ & $t_{(100)}^{z z}$ & $t_{(100)}^{x z}$ & $t_{(00\frac{1}{2})}^{x x}$ & $t_{(00\frac{1}{2})}^{z z}$ & $t_{(110)}^{x x}$ & $t_{(110)}^{z z}$ & $t_{(10\frac{1}{2})}^{x z}$ \\ \hline
		14.1 & 0.728 & 0.402 & -0.470 & -0.118 & 0.235 & 0.008 & -0.623 & 0.071 & -0.018 & -0.036 \\ 
		16.1 & 0.737 & 0.407 & -0.476 & -0.119 & 0.238 & 0.009 & -0.629 & 0.071 & -0.018 & -0.037 \\ 
		19.7 & 0.747 & 0.411 & -0.483 & -0.121 & 0.242 & 0.009 & -0.637 & 0.071 & -0.018 & -0.037 \\ 
		21.3 & 0.749 & 0.412 & -0.486 & -0.123 & 0.243 & 0.008 & -0.640 & 0.071 & -0.018 & -0.037 \\ 
		25.7 & 0.761 & 0.416 & -0.495 & -0.125 & 0.247 & 0.009 & -0.647 & 0.072 & -0.018 & -0.037 \\ 
		29.8 & 0.769 & 0.417 & -0.501 & -0.126 & 0.249 & 0.010 & -0.651 & 0.072 & -0.018 & -0.036 \\ 
		40.0 & 0.803 & 0.426 & -0.521 & -0.134 & 0.259 & 0.009 & -0.674 & 0.071 & -0.015 & -0.040 \\ 
		50.0 & 0.833 & 0.437 & -0.535 & -0.139 & 0.269 & 0.010 & -0.698 & 0.073 & -0.016 & -0.042 \\ 
		60.0 & 0.847 & 0.435 & -0.552 & -0.145 & 0.273 & 0.011 & -0.703 & 0.075 & -0.016 & -0.040 \\ 
		70.0 & 0.871 & 0.447 & -0.566 & -0.149 & 0.283 & 0.010 & -0.723 & 0.073 & -0.017 & -0.041 \\ 
		80.0 & 0.896 & 0.453 & -0.580 & -0.153 & 0.287 & 0.009 & -0.738 & 0.072 & -0.015 & -0.045 \\ 
		90.0 & 0.918 & 0.461 & -0.593 & -0.155 & 0.293 & 0.008 & -0.753 & 0.071 & -0.016 & -0.046 \\ 
		\hline
		\hline
		\label{T2}
	\end{tabular}
\end{table*}
By comparison, the difference between experimental and theoretical lattice values of La327 is less than (about) 1\% for $a$ ($c$, $l$), which is within a reasonable error. Besides, with the increase of pressure, the lattice parameters decrease monotonically. Additionally, we compare the electronic structures calculated using both experimental and theoretical lattice parameters shown in Fig.~S1 \cite{SM} (which also contains Refs. \cite{Kopietz__2010,Berges_PR_2002,Dupuis_PR_2021,Wang_PRB_2012,Xiang_PRB_2012,Wang_PRB_2013,Metzner_RMP_2012}) and find no significant difference, which confirms the reliability of theoretical calculation. Due to the fact that the lattice parameter are only measured up to 29.8 GPa experimentally, while we need to study the electronic structure and superconductivity up to 90 GPa, our calculations are based on the theoretically calculated lattice parameters in order to maintain consistency in our research. The band structures and DOS under different pressures are shown in Figs.~\ref{F1}(b) and \ref{F1}(c). As the pressure increases, it is found that the Fermi points at $E_\mathrm{F}$ remain almost unchanged, while the dispersion far from the $E_\mathrm{F}$ are significantly separated, enlarging the band width. Moreover, with increasing pressure, the DOS at the $E_\mathrm{F}$ first decreases and then increases slightly after 70 GPa, as shown in the subfigure in Fig.~\ref{F1}(c). Combining the orbital-projected band structure and partial DOS as shown in Figs.~S2(a) and S2(b) at 14.1 GPa \cite{SM}, it is evident that the states near the $E_\mathrm{F}$ are dominated by nickel $3d_{3z^2-r^2}$ and $3d_{x^2-y^2}$ orbitals, which is consistent with previous studies on La327 \cite{t-1,t-3,2023Nature,t-5,t-6,327arxivhu2}. In Fig.~S2(c) \cite{SM}, the three-dimensional Fermi surface is displayed, which exhibits two-dimensional-like feature.

Subsequently, the maximally localized Wannier functions \cite{Wannier1,Wannier2} of La327 with $3d_{3z^2-r^2}$ and $3d_{x^2-y^2}$ orbitals on each Ni atom are extracted by the Wannier90 code \cite{Wannier3} to construct a bilayer two-orbital tight-binding model
\begin{equation}
H_0=\sum_{i \delta, a b, \sigma} t_\delta^{a b} c_{i a \sigma}^{\dagger} c_{i+\delta b \sigma}+\sum_{i a \sigma} \varepsilon_a c_{i a \sigma}^{\dagger} c_{i a \sigma},
\end{equation}
where $t_\delta^{a b}$ is the hopping matrix element between the $a$ orbital on site $i$ and the $b$ orbital on site $i$ + $\delta$, $\sigma$ represents spin, and $\varepsilon_a$ is the on-site energy of the $a$-orbital. For simplicity, $x$ and $z$ represent the $3d_{x^2-y^2}$  and $3d_{3z^2-r^2}$ orbitals in the following, respectively. The good overlapping between DFT and Wannier band structures is also displayed in Fig.~S3 \cite{SM}, indicating the high quality of Wannier functions. The on-site energy and hopping parameters are listed in Table~\ref{T2}.  The vertical inter-layer hopping between the $z$-orbitals $t_{(00\frac12)}^{z z}$ is found to be the strongest, consistent with previous studies in La327 \cite{t-1,t-5,t-6}. 

\begin{figure}
	\includegraphics[width=\linewidth]{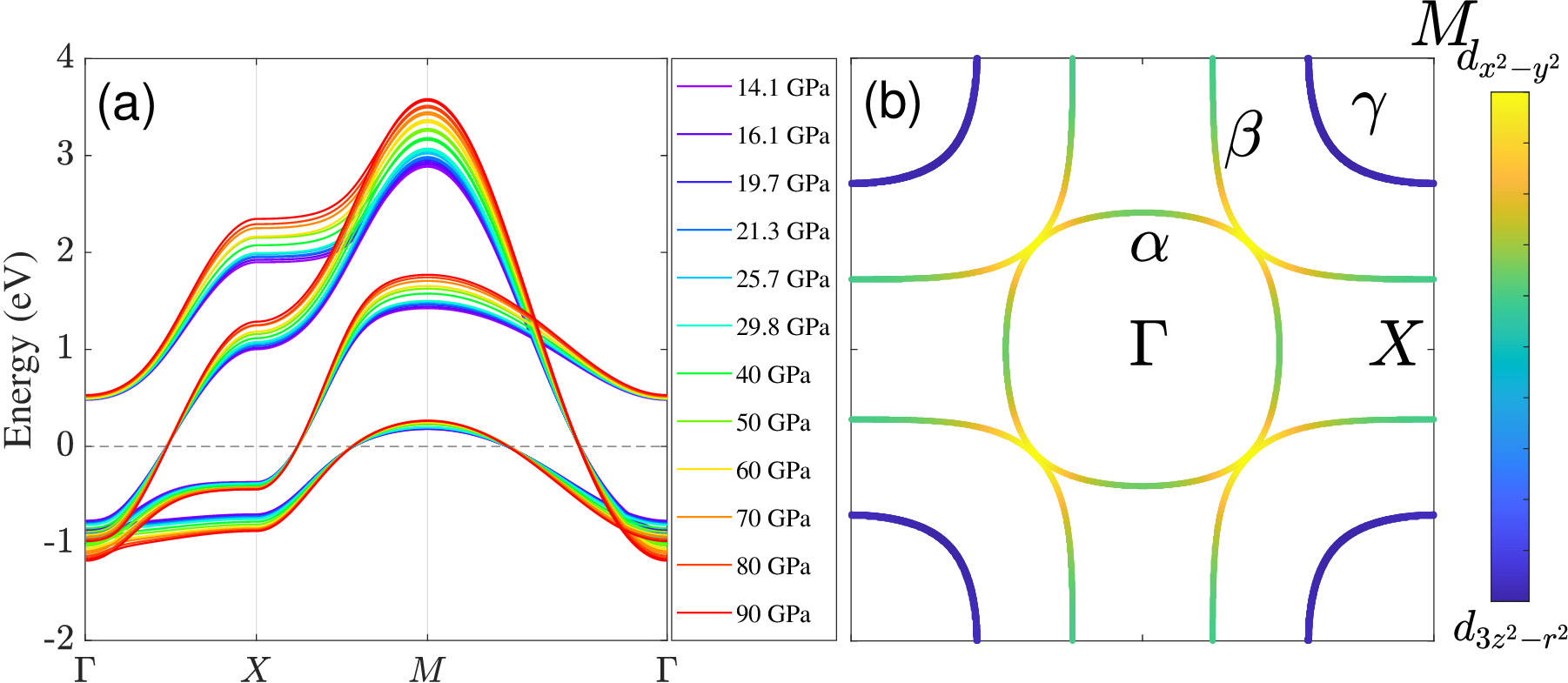}
	\caption{(a) Tight-binding band structures of La327 at pressures listed in Table~\ref{T2}. (b) Fermi surfaces obtained from the parameters shown at the pressure 14.1 GPa. The orbital weights of $d_{x^2-y^2}$ and $d_{3z^2-r^2}$ are represented by colors directly.}
	\label{fig:tb}
\end{figure}

In Fig.~\ref{fig:tb}(a), we plot the band structures of the tight-binding parameters at each pressure listed in Table~\ref{T2}. They reproduce nicely the DFT results in Fig.~\ref{F1}(b). The splitting of the band dispersion can be understood as follows. The Hamiltonian can be expressed as a block-diagonal form in the bonding and anti-bonding orbital basis, which hence, gives a perfect diagonal matrix with decoupled $x$ and $z$ orbital bands along the $M$-$\Gamma$ line due to the loss of the inter-orbital hopping \cite{t-1}.
Specifically at high symmetry points $M$ and $\Gamma$, we have the $x/z$ orbital bonding ($+$) and anti-bonding ($-$) energies $E^{x/z}_{M,\pm}= -4t^{xx/zz}_{100}+4t^{xx/zz}_{110}+\epsilon_{x/z}\pm t^{xx/zz}_{00\frac{1}{2}}$ and $E^{x/z}_{\Gamma,\pm}= 4t^{xx/zz}_{100}+4t^{xx/zz}_{110}+\epsilon_{x/z}\pm t^{xx/zz}_{00\frac{1}{2}}$. 
According to Table~\ref{T2} and Fig.~S5 \cite{SM}, these energies monotonically change with the pressure, which leads to the dispersion splitting.
On the other hand, the barely changed Fermi points in the dispersion implies the Fermi pockets are not changed under pressures. This is indeed the case from our explicit calculation of the Fermi pockets and the orbital contents. This can be explained by the concurrent change of the tight-binding parameters under pressure, see Fig.~S6 \cite{SM}. One typical result at pressure 14.1 GPa is shown in Fig.~\ref{fig:tb}(b). Similarly to the previous theoretical works \cite{t-1,t-2}, there are three Fermi pockets $\alpha$, $\beta$ and $\gamma$. The $\alpha$ and $\beta$ pockets have mixed orbital weights (except in the diagonal direction), while the $\gamma$ pocket is dominated by the $d_{3z^2-r^2}$ orbital.

{\it FRG analysis}.
\begin{figure}
	\includegraphics[width=\linewidth]{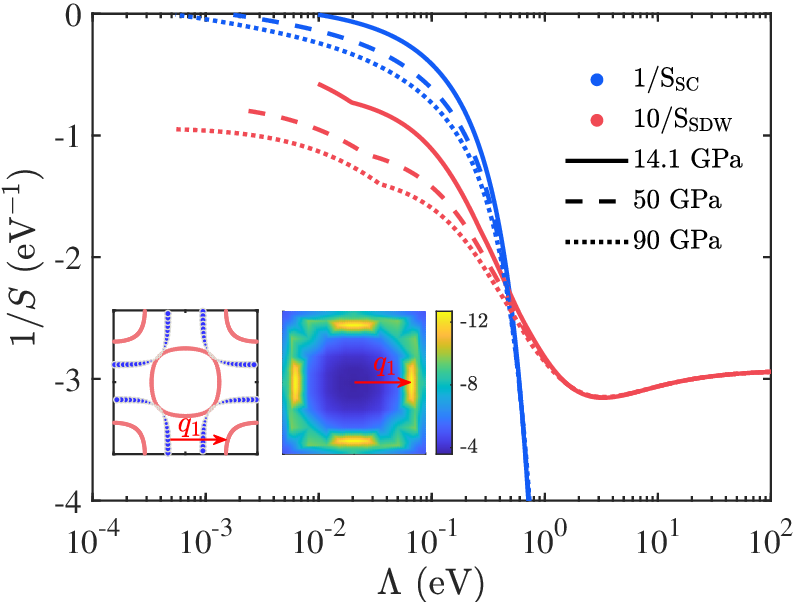}
	\caption{FRG flows of $S^{-1}$ versus $\Lambda$ in the SC and SDW channels of La327, respectively, at pressures 14.1, 50, and 90 GPa. The left subfigure present the gap function on the Fermi surfaces. The right subfigure present the leading negative $S(\0q)$ in the SDW channel. Both subfigures are the results at pressure 50 GPa.}
	\label{fig:frg}
\end{figure}
The well-constructed tight-binding models discussed above enables us to study the superconductivity driven by correlation effects in La327 system under pressure. In this Letter, we use SM-FRG to calculate the one-particle-irreducible four-point vertex $\Gamma_{\Lambda}$ iteratively against a running infrared cutoff energy scale $\Lambda$. The same vertex $\Gamma_{\Lambda}$ is re-expressed as scattering matrices between fermion bilinears in the superconducting (SC), spin-density-wave (SDW) and charge-density-wave (CDW) channels. We monitor the negative leading eigenvalue $S$ of these scattering matrices during the FRG flow. The first divergence of the $S$ (among the three channels) at the scale $\Lambda_c$ signals an emerging order described by the scattering channel, the scattering momentum, and the eigen scattering mode. 
In addition to the fact that the SM-FRG treats all possible orderings on equal footing, it has the further advantage to obtain directly the transition temperature $T_c\sim \Lambda_c$. More technical details can be found in Refs.~\cite{Wang_PRB_2012,Wang_PRB_2013,Tang_PRB_2019,t-2,Yang_prb_2024} and the Supplemental Material \cite{SM}. 
In the following, we first show three typical FRG results at different cases, followed by the discussion on the systematic pressure dependence of $T_c$. 

\begin{figure}
	\includegraphics[width=\linewidth]{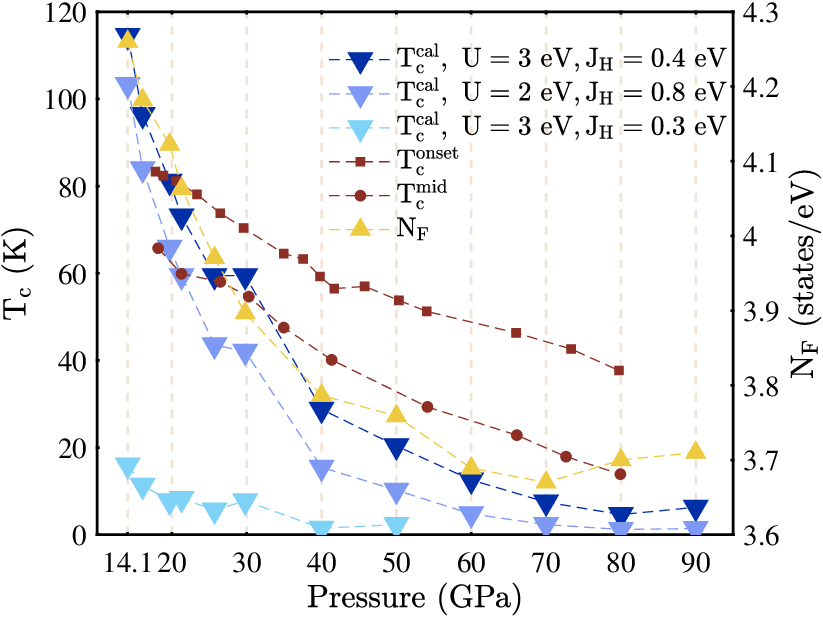}
	\caption{Phase diagram of superconducting $T_c$ versus pressure of La327. The $T^{onset}_c$ and $T^{mid}_c$ are extracted from the experimental work \cite{327pressure} for comparison.}
	\label{fig:phase}
\end{figure}
	
In Fig.~\ref{fig:frg}, we plot the inverse of the values $S$ versus $\Lambda$ in the SC and SDW channels, respectively, at pressures 14.1, 50, and 90 GPa.
The CDW channel is much weaker and omitted for clarity. We choose the intraorbital Hubbard repulsion $U=3$ eV and Hund's coupling $J_H=0.4$ eV. The Kanamori relations are also used, so that we have the interorbital repulsion $U'=U-2J_H$ and inter-orbital pair-hopping $J_P=J_H$.    
The initial FRG flows at higher energy scales are quite similar for the three different pressures, where the bare interactions dominate and hence the SDW channel is stronger. At low energy scales, however, the SDW channel becomes subleading and turns to saturate because of the lack of phase space for low-energy particle-hole excitations. The momentum dependent eigenvalue in the SDW channel is presented in the right inset, showing peaks in the (1,0) and (0,1) directions. 
At intermediate energy scales where the SDW channel grows, the SC channel begins to be induced through channel overlap. The latter then eventually diverges by Cooper mechanism. This is a typical manifestation of spin fluctuation mediated superconductivity. Similar results are obtained for all cases listed in Table~\ref{T2}. From the eigen scattering mode in the SC channel, we obtain the pairing gap function on the Fermi pockets (left inset), showing  $s_\pm$-wave symmetry. Up to $C_{4v}$ symmetry, there exists a momentum $\mathbf{q_1}$ that connects two Fermi pockets with opposite pairing signs. This particular momentum is just the peak momentum in the SDW channel (right inset), which reconfirms the connection between pairing and spin fluctuations here. 

We see from Fig.~\ref{fig:frg} that the critical energy scale $\Lambda_c$ drastically drops as the pressure increases from 14.1 to 50 GPa and 90 GPa, and this can be clearly related to the weakening of spin fluctuations (at the final stage of FRG flow). This implies the high sensitivity of the transition temperature $T_c\sim \Lambda_c$ on the pressure. We therefore perform systematic SM-FRG calculations for a sequence of pressures listed in Table~\ref{T2} to obtain the pressure dependence of the transition temperature $T^{cal}_c$ shown in Fig.~\ref{fig:phase} (down triangles). The experimental data of superconducting $T^{onset}_c$ (filled squares) and $T^{mid}_c$ (filled circles) \cite{327pressure} are also displayed for comparison. Remarkably, there is a notable consistency between the calculated results and the experimental observations.    
Moreover, to see the robustness of our SM-FRG results, we have performed systematic calculations by varying $U$ and $J_H$. The results are qualitatively similar, see Fig.~\ref{fig:phase}, except that the lowest $J_H$ yields the lowest $T_c$, implying the importance of Hund's coupling. We also note that while the normal state DOS (yellow triangles) is higher at 80 GPa than that at 70 GPa, our theoretical $T_c$ does not follow the trend in DOS, but is in agreement with the experiment. This indicates the necessity of treating the band structure and the correlations systematically, as we do in our FRG calculations. 

{\it Conclusion}. 
Combing the DFT and SM-FRG calculations, the electronic structures, tight-binding model, pairing mechanism, and pairing symmetry are systematically investigated under different pressures in the superconducting regime of La327. We find the pairing symmetry is consistently $s_\pm$-wave, triggered by spin fluctuations which become increasingly weakened by pressure and consequently lead to decreasing superconducting transition temperature, in qualitative agreement with the experiment. We argue that the pressure dependence is difficult, if not impossible, to be explained by the local-moment picture of the electrons, while the success in our theory sheds light on the importance of the itineracy of electrons in La327, at least in the superconducting regime. 

{\it Acknowledgments}. 
We thank Meng Wang for experimental data and discussion. This work is supported by National Natural Science Foundation of China (Grant No. 12074213, No. 12374147, No. 12274205, No. 92365203, No. 11874205, and No. 11574108), National Key R\&D Program of China (Grant No. 2022YFA1403201), and Major Basic Program of Natural Science Foundation of Shandong Province (Grant No. ZR2021ZD01).


\bibliography{327pressure}

\end{document}